\documentclass[a4paper,fleqn,usenatbib]{mnras}
\usepackage{longtable}
\usepackage{lscape}
\usepackage{subfig}
\usepackage{hyperref}

\usepackage[T1]{fontenc}
\usepackage{ae,aecompl}

\usepackage{graphicx}	
\usepackage{amsmath}	
\usepackage{amssymb}	

\title[Halo transitional brown dwarfs]{Primeval very  low-mass stars and brown dwarfs -- III. The halo transitional brown dwarfs }

\author[Z. H. Zhang et al.]{Z. H. Zhang,$^{1,2,3}$\thanks{E-mail:
zenghuazhang@hotmail.com}\thanks{PSL Fellow}
 D. J. Pinfield,$^{4}$ M. C. G\'{a}lvez-Ortiz,$^{5}$ D. Homeier,$^{6}$ 
 \newauthor 
A. J. Burgasser,$^{7}$  N. Lodieu,$^{2,3}$ E. L. Mart{\'i}n,$^{5}$ M. R. Zapatero Osorio,$^{5}$  F. Allard,$^{8}$ 
\newauthor 
H. R. A. Jones,$^{4}$   R. L. Smart,$^{9}$    B. L{\'o}pez Mart{\'i},$^{10}$ B. Burningham$^{4}$ and R. Rebolo$^{2,3}$
\\
$^{1}$GEPI, Observatoire de Paris, Universit{\'e} PSL, CNRS, 5 Place Jules Janssen, 92190 Meudon, France \\
$^{2}$Instituto de Astrof{\'i}sica de Canarias, E-38205 La Laguna, Tenerife, Spain \\
$^{3}$Universidad de La Laguna, Dept. Astrof{\'i}sica, E-38206 La Laguna, Tenerife, Spain \\
$^{4}$Centre for Astrophysics Research, Science and Technology Research Institute, University of Hertfordshire, Hatfield AL10 9AB, UK \\
$^{5}$Centro de Astrobiolog{\'i}a (CSIC-INTA), Ctra. Ajalvir km 4, E-28850 Torrej{\'o}n de Ardoz, Madrid, Spain \\
$^{6}$Zentrum f{\"u}r Astronomie der Universit{\"a}t Heidelberg, Landessternwarte, K{\"o}nigstuhl 12, D-69117 Heidelberg, Germany  \\
$^{7}$Center for Astrophysics and Space Science, University of California San Diego, La Jolla, CA 92093, USA \\
$^{8}$Univ Lyon, ENS de Lyon, Univ Lyon 1, CNRS, Centre de Recherche Astrophysique de Lyon, UMR5574, F-69007, Lyon, France \\
$^{9}$Istituto Nazionale di Astrofisica, Osservatorio Astronomico di Torino, Strada Osservatrio 20, I-10025 Pino Torinese, Italy  \\
$^{10}$Saint Louis University - Madrid Campus, Avenida del Valle 34, E-28003 Madrid, Spain
}

\date{Accepted 2018 May 21. Received 2018 May 18; in original form 2017 September 7}

\begin{document}

\label{firstpage}
\pagerange{\pageref{firstpage}--\pageref{lastpage}}
\maketitle
\begin{abstract}
We report the discovery of an esdL3 subdwarf, ULAS J020858.62+020657.0, and a usdL4.5 subdwarf, ULAS J230711.01+014447.1. They were identified as L subdwarfs by optical spectra obtained with the Gran Telescopio Canarias, and followed up by optical-to-near-infrared spectroscopy with the Very Large Telescope. We also obtained an optical-to-near-infrared spectrum of a previously known L subdwarf, ULAS J135058.85+081506.8, and reclassified it as a usdL3 subdwarf. These three objects all have typical halo kinematics. They have $T_{\rm eff}$ around 2050--2250 K, $-$1.8 $\leq$ [Fe/H] $\leq -$1.5, and mass around 0.0822--0.0833 M$_{\odot}$, according to model spectral fitting and evolutionary models. These sources are likely halo transitional brown dwarfs  with unsteady hydrogen fusion, as their masses are just below the hydrogen-burning minimum mass, which is $\sim$ 0.0845 M$_{\odot}$ at [Fe/H] = $-$1.6 and $\sim$ 0.0855 M$_{\odot}$ at [Fe/H] = $-$1.8. Including these, there are now nine objects in the `halo brown dwarf transition zone', which is a `substellar subdwarf gap' that spans a wide temperature range within a narrow mass range of the substellar population.
\end{abstract}




\section{Introduction}
The formation of brown dwarfs that have masses below the hydrogen-burning minimum mass (HBMM) was predicted over a half century ago \citep{kuma63,haya63}. Modern evolutionary models indicate that the HBMMs are between 0.07 and 0.092 M$_{\odot}$ from solar to primordial metallicity \citep{bara15,burr01}. One of the motivations to find these brown dwarfs was the potential of a new celestial population with masses between stars and planets to test stellar/substellar formation theory. 

A few thousand ultracool dwarfs have been identified spectroscopically since the 1990s. To better understand the physics of stellar nuclear ignition and the inner structure of objects across the stellar--substellar boundary \citep{cha97}, it is crucial to distinguish between stars and brown dwarfs with mass around the HBMM. However, evolution leads to mass/age degeneracy when considering the luminosity and effective temperature ($T_{\rm eff}$) of very low-mass star and brown dwarf populations. Therefore, it takes a lot of efforts to assess the substellar status of field ultracool dwarfs near the HBMM \citep[e.g.][]{diet14,dupu17}.  

Brown dwarfs in the thick disc and halo have lower level of mass/age degeneracy than those in the thin disc, as they have similar ages to the stellar populations in the thick disc \citep[$\sim$ 8--10 Gyr; ][]{red06,kili17} and the halo \citep[$\sim$ 10--13 Gyr; ][]{dott10,jofr11}, and they have evolved to a stage with slowed-down of cooling. However, they are distributed in a wide range of metallicity and have different HBMMs at different metallicities. Therefore, the metallicity measurements offer significant scope to tackle the mass/metallicity degeneracy problem for thick disc and halo brown dwarfs. 

In the first paper of a series under the title {\sl `Primeval very low-mass stars and brown dwarfs'}, we defined a new classification scheme for L subdwarfs and characterized a sample of 22 metal-deficient ultracool subdwarfs \citep[][hereafter Paper I]{zha17a}. In the second paper of the series we assessed the stellar--substellar boundary in the $T_{\rm eff}$ versus [Fe/H] plane based on theoretical model prediction of the HBMM \citep[][hereafter Paper II]{zha17b}. This provided an alternative method to assess substellar status for nearby ultracool subdwarfs, and revealed a `halo brown dwarf transition zone' representing a narrow mass range in which unsteady nuclear fusion occurs -- a `substellar subdwarf gap' for mid L to early T types.  
Currently, there are six L subdwarfs known in this transition zone: SDSS J010448.46+153501.8  (\citealt{lod17}; Paper II), 2MASS J05325346+8246465 \citep{bur03}, 2MASS J06164006$-$6407194  \citep[2M0616;][]{cus09}, SDSS J125637.13$-$022452.4  \citep[SD1256;][]{siv09}, ULAS J151913.03$-$000030.0 (UL1519; Paper I), and 2MASS J16262034+3925190 \citep[2M1626;][]{bur04}. 

This is the third paper of the series following Paper I and Paper II, in which we grow the halo transitional brown dwarf population further, presenting the discovery of three new members. 
The observations are presented in Section \ref{sobs}. Section \ref{spro} presents constraints on the physical characteristics of these objects. Finally Section \ref{ssac} presents a summary and conclusions.

\begin{table}
 \centering
  \caption[]{Properties of UL0208, UL1350, and UL2307.}
\label{prop}
  \begin{tabular}{l c c c}
\hline
Parameter & UL0208 & UL1350 & UL2307  \\	
\hline 
 $\alpha$ (J2000) & $02^{\rm h}08^{\rm m}58\fs62$ & $13^{\rm h}50^{\rm m}58\fs85$ & $23^{\rm h}07^{\rm m}11\fs01$ \\
 $\delta$ (J2000) & $+02\degr06\arcmin57\farcs0$ & $+08\degr15\arcmin06\farcs8$ & $+01\degr44\arcmin47\farcs1$   \\
Epoch & 2010-08-28 & 2006-07-08 & 2009-08-11 \\
SDSS $i$  & 21.54$\pm$0.07&  21.22$\pm$0.08 & 22.53$\pm$0.22 \\
SDSS $z$ &19.86$\pm$0.07 & 19.47$\pm$0.06 & 19.91$\pm$0.09 \\
PS1 $i$ & 21.39$\pm$0.03 & 21.14$\pm$0.07 & 21.69$\pm$0.16 \\
PS1 $z$ & 19.99$\pm$0.03 & 19.68$\pm$0.05 & 20.16$\pm$0.04 \\
PS1 $y$ & 19.50$\pm$0.06 & 19.32$\pm$0.04 & 19.57$\pm$0.08 \\
UKIDSS $Y$  & 18.76$\pm$0.05& 18.66$\pm$0.05 &  18.99$\pm$0.08 \\
UKIDSS $J$  & 18.00$\pm$0.04& 17.93$\pm$0.04 & 18.15$\pm$0.06 \\
UKIDSS $H$ & 17.88$\pm$0.13 & 18.07$\pm$0.10 & 18.34$\pm$0.12 \\
UKIDSS $K$ & 17.62$\pm$0.16 & 17.95$\pm$0.15 & 18.17$\pm$0.18 \\
Spectral type & esdL3$\pm$1 & usdL3$\pm$1 & usdL4.5$\pm$1 \\
Distance (pc) & 171$^{+32}_{-27}$ & 176$^{+25}_{-22}$ & 157$^{+19}_{-18}$  \\
$\mu_{\rm RA}$(mas/yr) &  147.1$\pm$8.4 & $-$252.8$\pm$5.4   & $-$33.0$\pm$9.1 \\
$\mu_{\rm Dec}$(mas/yr) &  $-$143.1$\pm$10.4  & $-$248.5$\pm$4.3 & $-$202.6$\pm$2.6  \\
$V_{\rm tan}$ (km/s) & 166$\pm$31 & 239$\pm$39 & 205$\pm$30   \\
RV (km/s) & 52$\pm$10 & 58$\pm$9 &  $-$215$\pm$11  \\
$U$ (km/s) & $-$58$\pm$27 & $-$23$\pm$26& 67$\pm$41 \\
$V$ (km/s) &  $-$151$\pm$50 & $-$299$\pm$88 & $-$232$\pm$48 \\
$W$ (km/s) &  $-$64$\pm$28  & 33$\pm$28 &  106$\pm$40 \\
$T_{\rm eff}$ (K) & 2250$\pm$120  & 2250$\pm$120 & 2050$\pm$120  \\
$\rm {[Fe/H]}$ & $-$1.5$\pm$0.2 & $-$1.8$\pm$0.2 & $-$1.7$\pm$0.2 \\
$M$ (\%M$_{\odot}$) & 8.27$\pm$0.15 & 8.33$\pm$0.15 & 8.22$\pm$0.15  \\\hline
\end{tabular}
\end{table}

\section{Observations}
\label{sobs}
ULAS J020858.62+020657.0 (UL0208) and ULAS J230711.01+014447.1 (UL2307) were selected as L subdwarf candidates from the UKIRT Infrared Deep Sky Survey's \citep[UKIDSS; ][]{law07} Large Area Survey (LAS) and the Sloan Digital Sky Survey \citep[SDSS; ][]{yor00}. The selection criteria and procedure are described in Paper I.  The properties of UL0208 and UL2307 are summarized in Table \ref{prop}. The spectroscopic observations are summarized in Table \ref{tsdlob}.
 
\subsection{GTC spectroscopy}
UL0208 and UL2307 were first identified as L subdwarfs by their optical spectra obtained with the Optical System for Imaging and low Resolution Integrated Spectroscopy \citep[OSIRIS;][]{cepa00} instrument on the Gran Telescopio Canarias (GTC). An R500R grism and 0.8 arcsec slit were used for observations of UL0208 and UL2307, which provide a resolving power of $\sim$ 500 and cover a wavelength range of 480--1020 nm. These spectra were reduced using standard procedures within {\scriptsize  IRAF}\footnote{IRAF is distributed by the National Optical Observatory, which is operated by the Association of Universities for Research in Astronomy, Inc., under contract with the National Science Foundation.}. Standard stars G158--100 \citep[dG-K;][]{oke90} and GD140 \citep[DA WD;][]{mass88} were used for the flux calibration of UL0208 and UL2307, respectively. The contamination from the second-order filter is not corrected for standard stars; therefore, the flux calibration beyond 900 nm is  based on the extension of the first-order response function and somewhat uncertain.  However, the flux of these OSIRIS spectra are roughly consistent with their X-shooter spectra (we obtained later) at 900--1020 nm except the telluric region at 925--965 nm. Telluric is not corrected for these OSIRIS spectra. The signal-to-noise ratio (SNR) at 850 nm is $\sim$60 and $\sim$55 for the spectra of UL0208 and UL2307, respectively. 

\subsection{VLT spectroscopy}
UL0208 and UL2307 were followed up with the X-shooter spectrograph \citep{ver11} on the Very Large Telescope (VLT) of the European Southern Observatory (ESO), together with another previously known L subdwarf, ULAS J135058.85+081506.8 \citep[UL1350;][]{lod10,lod17}. We also observed an L0.5 dwarf radial velocity (RV) standard, DENIS-P J144137.3$-$094559 \citep[DE1441;][]{mart99,bail04} with X-shooter for RV measurements of our objects. 
 All X-shooter spectra were observed in an ABBA nodding mode with a 1.2 arcsec slit providing a resolving power of 6700 in the visible (VIS) arm (530--1020 nm) and 4000 in the near-infrared (NIR) arm (990--2480 nm). These spectra were reduced to flux- and wavelength-calibrated 2D spectra with ESO Reflex \citep{freu13}. The 1D spectra were extracted from these 2D spectra using the {\scriptsize IRAF} task {\scriptsize APSUM}.
Telluric corrections were achieved using telluric standards which were observed right after or before our targets at a very close airmass. 

The spectrum of UL0208 has an SNR (per pixel) of $\sim$3 at 830 nm and $\sim$6 at 1300 nm. The spectrum of UL2307 has an SNR of $\sim$5 at 830  and 1300 nm. The spectrum of UL1350 has an SNR of $\sim$6 at 830 and 1300 nm. These spectra presented in this paper have a resolving power of $\sim$600 as they were smoothed by 101 pixels and 51 pixels for the VIS and NIR arms to increase the SNR by factors of $\sim$10 and $\sim$7, respectively. The spectrum of the  DE1441 has an SNR of $\sim$ 20 at 830 nm and $\sim$ 70 at 1300 nm.

\begin{table*}
 \centering
  \caption[]{Summary of the spectroscopic observations.   } 
\label{tsdlob}
  \begin{tabular}{c c c c c c c c c c c }
\hline
    Name & Telescope & Instrument  & UT date & Seeing  & Airmass   &  $T_{\rm int}$(VIS) &  $T_{\rm int}$(NIR)  & Telluric & SpT & Airmass  \\
    & & & & (\arcsec) & & (s) & (s)  & star & \\
\hline
 UL0208 & GTC & OSIRIS  & 2013-09-09 & 0.80 & 1.76 &  2700 & --- & ---  & --- & ---  \\
 UL2307 & GTC & OSIRIS   & 2013-07-05 & 0.80 & 1.27 &  3600 & --- & --- & --- & ---   \\
 UL0208  &  VLT & X-shooter  & 2014-11-29 & 1.13 & 1.13  &  3384  &  3480 & HD 22686 & A0 V & 1.14 \\
UL2307   & VLT & X-shooter  & 2015-09-11 & 1.02 & 1.13   &  3420  &  3600 & HIP 105164 & B7 III & 1.09 \\
UL1350  & VLT & X-shooter  & 2016-03-22 & 0.59 & 1.20   & 3480  &  3600 & HIP 79439 & B9 V & 1.42 \\
DE1441 & VLT & X-shooter & 2016-03-19 & 0.66 & 1.11 &  1120  &  1200 & HIP 76836 &  B9.5 V & 1.11 \\
\hline
\end{tabular}
\end{table*}

\begin{figure*}
\begin{center}
 \includegraphics[width=\textwidth]{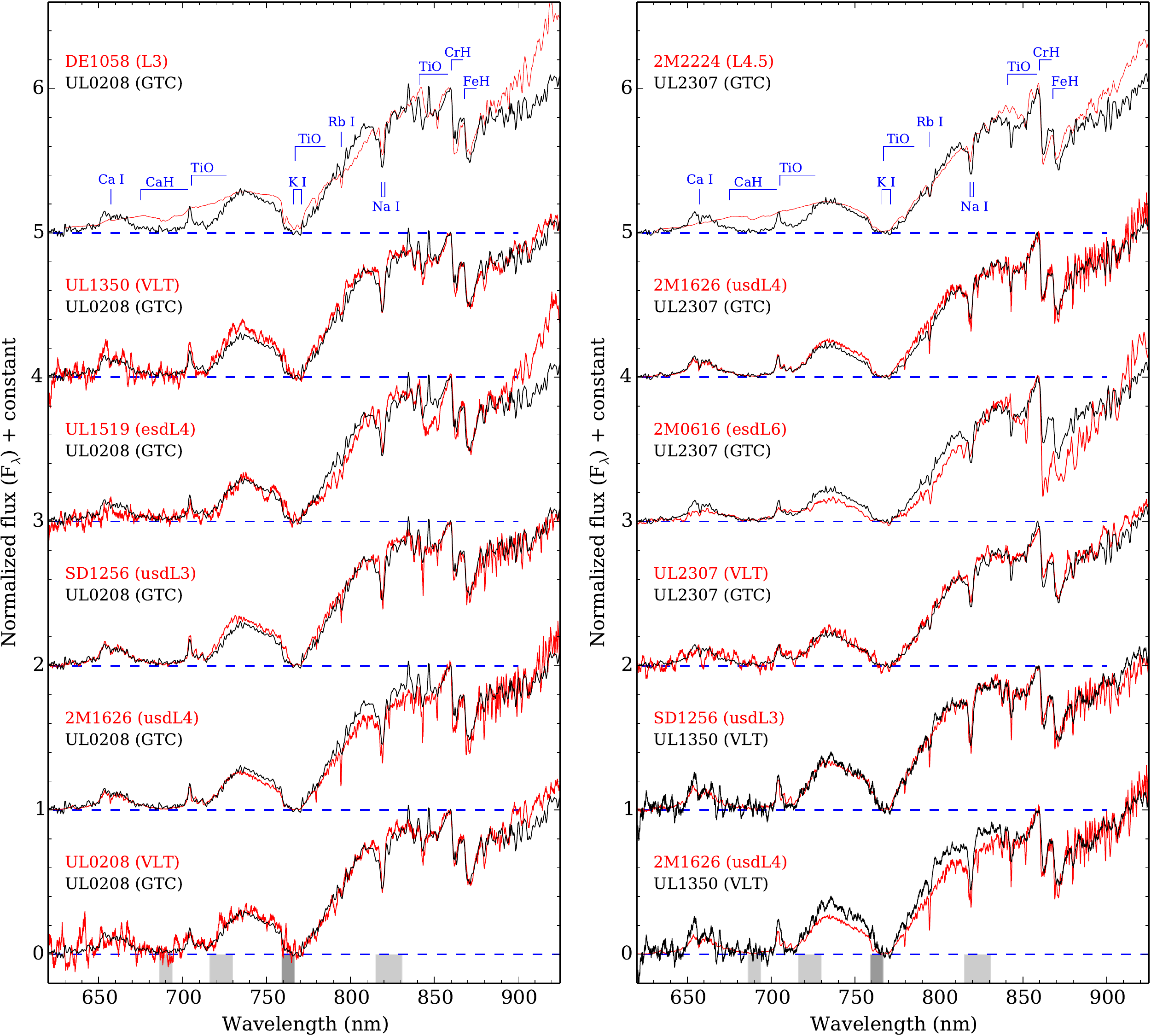}
\caption[]{The optical spectra of UL0208, UL1350, and UL2307, compared to spectra of known objects from \citet[][SD1256]{bur09}; \citet[][2M1626]{bur07};  Paper I (UL1519); \citet[][2M0616]{cus09}; \citet[][DE1058]{kirk99}; and \citet[][2M2224]{kirk00}. Spectra are normalised near 860 nm. Telluric absorption wavelengths are indicated with grey strips at the bottom and have been corrected for our VLT spectra, but not for our GTC spectra. Lighter and thicker shaded bands indicate regions with weaker and stronger telluric effects.}
\label{spvis}
\end{center}
\end{figure*}

\begin{figure}
\begin{center}
 \includegraphics[width=\columnwidth]{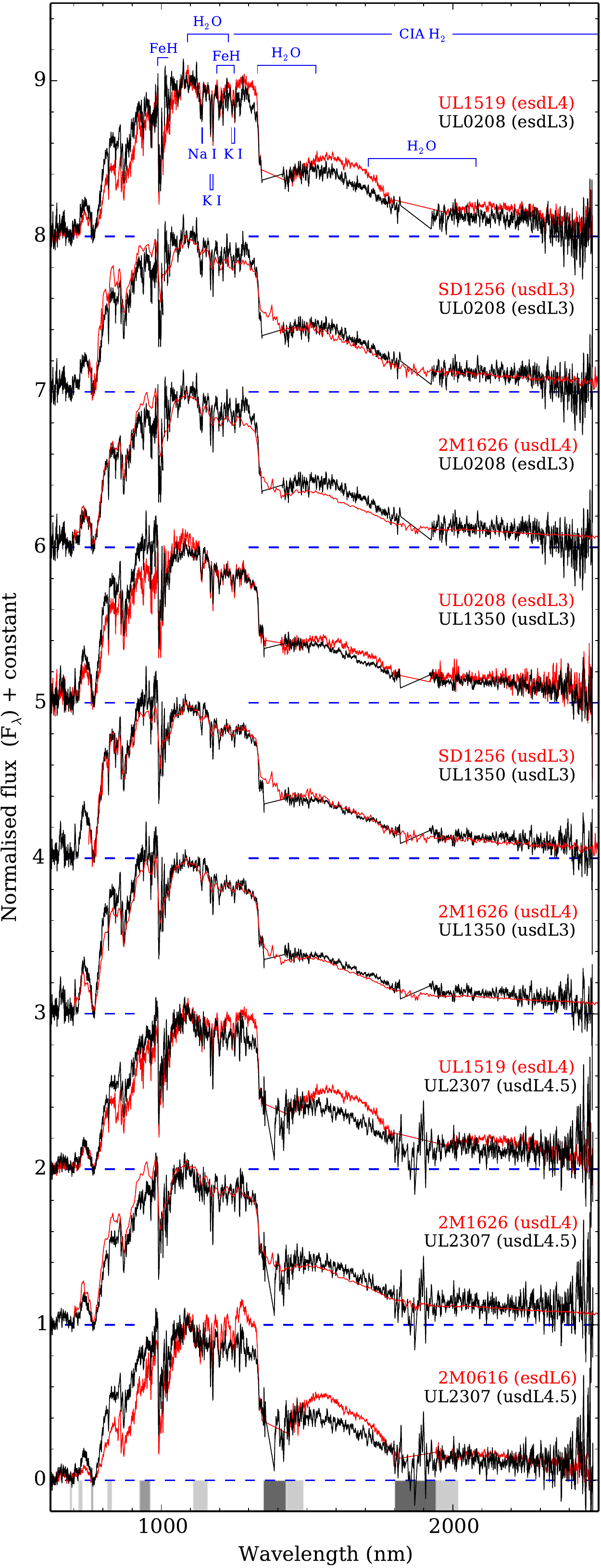}
\caption[]{The optical--NIR spectra of UL0208, UL1350, and UL2307 compared to spectra of known objects from \citet[][SD1256]{bur09}; \citet[][2M1626]{bur04}; and Paper I (UL1519 and 2M0616). Spectra are normalised near 1100 nm. }
\label{spnir}
\end{center}
\end{figure}

\section{Characteristics}
\label{spro}
\subsection{Spectral classification}
\label{sspec}
We classify UL0208, UL2307 and UL1350 following our L subdwarf classification scheme (Paper I). L subdwarfs are classified into three metallicity subclasses, subdwarf (sdL), extreme subdwarf (esdL), and ultra-subdwarf (usdL), based on their optical-to-NIR spectra. The classification scheme was set up such that the subclass sequence follows a decreasing trend in metallicity. The overall profiles of L subdwarfs respond to both $T_{\rm eff}$ and metallicity. The overall optical-to-infrared spectral energy distribution becomes bluer for both warmer $T_{\rm eff}$ and/or lower metallicity; however, features such as the 705--730 and 840--860 nm TiO bands are specifically metal-sensitive (see table 3 in Paper I).

The population of known L subdwarfs with good-quality optical-to-NIR spectra does not fully populate every subclass/subtype in the scheme, so we followed a two-stage approach. First we compared our target spectra to those of the available classified subdwarfs and identified objects that provided a reasonable match (in terms of overall morphology and important spectral features). Then we considered each match in more detail, and made decisions about optimal choice and any intermediate classifications.

There were four classified subdwarfs (with good optical-to-NIR spectra) that gave reasonable matches to our three objects. These were the esdL6  2M0616 \citep{cus09}, the usdL3 SD1256 \citep{siv09}, the esdL4 UL1519 (Paper I) and the usdL4 2M1626 \citep{bur04}. Figs \ref{spvis} and \ref{spnir} show the optical and NIR spectra of our objects (UL0208, UL2307, and UL1350) compared to those of the four classified L subdwarfs. 

As can be seen, UL0208 has a similar optical spectrum to the usdL3 SD1256, however, it has slightly less flux at 720--770 nm indicating a higher metallicity. Also, compared to the esdL UL1519, UL0208 has slightly more flux at 780--810 nm, weaker TiO absorption at around 840--860 nm, and stronger NIR suppression, indicating a slightly lower metallicity than UL1519. Comparison between the optical spectra of UL0208 and 2M1626 shows that UL0208 has an earlier spectral type (SpT) than 2M1626. Therefore we classify UL0208 as an esdL3 subdwarf.

The optical spectrum of UL2307 appears to be intermediate between those of the usdL4 2M1626 and the esdL6 2M0616 (see Fig. \ref{spvis}). However, its NIR spectrum is significantly more suppressed than that of these subdwarfs, with an overall optical-to-NIR morphology that is closer to 2M1626 than to 2M0616. We therefore assign UL2307 the same subclass as 2M1626. Having said this, UL2307 has slightly less flux in the optical and slightly more flux in the NIR compared to 2M1626, which leads us to assign it a slightly later subtype. We classified UL2307 as a usdL4.5 subdwarf. 

Fig. \ref{spvis} shows that both UL0208 and Ul2307 have very different optical spectral features (e.g. TiO) from the standard L3 dwarf DENIS-P J1058.7$-$1548 \citep[DE1058;][]{delf97,kirk99} and L4.5 dwarf 2MASS J2224438$-$015853 \citep[2M2224;][]{kirk00} because of their much lower metallicity. Metallicity has a stronger influence on the spectral profile of L subdwarfs in the NIR than in the optical (e.g. fig. 9 in Paper I).

UL1350 was originally classified as sdL5 by \citet{lod10} based on a low SNR optical spectrum, but was then re-classified sdL3.5--4 based on an improved SNR optical spectrum \citep{lod17}. In Paper I, we suggested that UL1350 should have the same spectral type as SD1256 (usdL3). And here we are now able to assess the full optical-NIR spectra. These spectra (shown in Figures \ref{spvis} and \ref{spnir}) show that UL1350 has an optical-to-NIR morphology that is very similar to SD1256, and we thus classify UL1350 as usdL3.

\subsection{Halo kinematics}
\label{shk}
We derived spectroscopic distance estimates for UL0208, UL1350, and UL2307 using the relationship between spectral type and $J$- and $H$-band absolute magnitude shown in fig. 16 of Paper I. We computed distances using both $J$ and $H$ band magnitudes which for all three sources agreed within the uncertainties; the average distances are listed in Table \ref{prop}. Upper and lower limits were determined by considering the $J$ and $H$ band constraints, and positive and negative error bars were assigned for each object.

Proper motions were determined using multi-epoch optical and NIR data. UL0208, UL1350, and UL2307 were observed by SDSS on 2009 September 27, 2003 April 27 and 2008 October 02, respectively. They were subsequently observed by UKIDSS and Pan-STARRS1 \citep[PS1;][]{cham16}. Multi-epoch PS1 images of each target were combined, with mean epochs of 2012 July 31, 2011 March 12, and 2012 October 23, for UL0208, UL1350, and UL2307, respectively. The proper motions of UL0208 and UL1350 were measured using SDSS $z$- and PS1 $z$-band images which had baselines of 2.84 and 7.88 yr respectively, following the method described in section 3.2 of Paper II. The proper motion of UL2307 was measured from PS1 $z$- and UKIDSS $J$-band images which provided a baseline of 3.20 yr. The numbers of reference stars used for proper motion measurements were 22, 9 and 9 for UL0208, UL1350, and UL2307, respectively. Our measured proper motions are listed in Table \ref{prop}.

We measured RVs for UL0208, UL2307 and UL1350 by cross correlation of their X-shooter optical and NIR spectra to that of an DE1441 which has a RV of $-27.9\pm1.2$ km s$^{-1}$ \citep{bail04}. We smoothed our original X-shooter spectra by 21 pixels in the VIS and 11 pixels in the NIR to increase the SNR. The RV differences between our objects and the standard were measured with  the {\scriptsize IRAF} task {\scriptsize FXCOR}. Then the RVs were corrected for the barycentric effects. The final RVs of UL0208, UL2307 and UL1350 are listed in Table \ref{prop}. 

The Galactic $UVW$ space motions of UL0208, UL2307 and UL1350 were determined using our spectroscopic distances, RVs and proper motions following \citet{clar10}. The resulting $U, V$ and $W$ velocities are listed in Table \ref{prop}. 
UL0208, UL2307 and UL1350 have typical halo kinematics (see fig. 17 of Paper I), in keeping with their low metallicity making them halo members with ages of $\sim$ 10--13 Gyr \citep[][]{dott10,jofr11}. 

 We calculated the halo membership probabilities of these three objects based on kinematics and population fractions of thin disc (0.93), thick disc (0.07) and halo (0.006) stars in the solar neighbourhood \citep{red06}. UL1350 and UL2307 have halo membership probabilities of 99.9\% and 99.1\%, respectively.  UL0208 has a higher thick disc probability (82.9\%) and a lower halo probability (17.1\%), as the thick disc fraction is $\sim$ 10 times higher than the halo fraction. UL0208 would have a halo probability of 70.1\% if we applied an equal fraction of halo and thick disc populations in the calculation. The halo membership of UL0208 becomes robust when considering its typical halo metallicity of [Fe/H] = $-$1.5.

\subsection{Atmospheric properties}
\label{sap}

We used the BT-Dusty models \citep{alla14} to constrain the atmospheric parameters of UL0208, UL1350, and UL2307. The model grids we used cover 1900 K $\leq T_{\rm eff} \leq$ 3500 K, $-2.5 \leq$ [Fe/H] $\leq -1.0$ and 5.0 $\leq$ log $g$ $\leq$ 5.5, with intervals of 100 K for $T_{\rm eff}$, 0.5 dex for [Fe/H], and 0.5 dex for log $g$. Linear interpolation between some model spectra was used if this yielded a significantly improved fit. We followed the procedure described in section 3.3 of Paper II to find best-fitting model spectra to our sources. 

\begin{figure}
\begin{center}
 \includegraphics[width=\columnwidth]{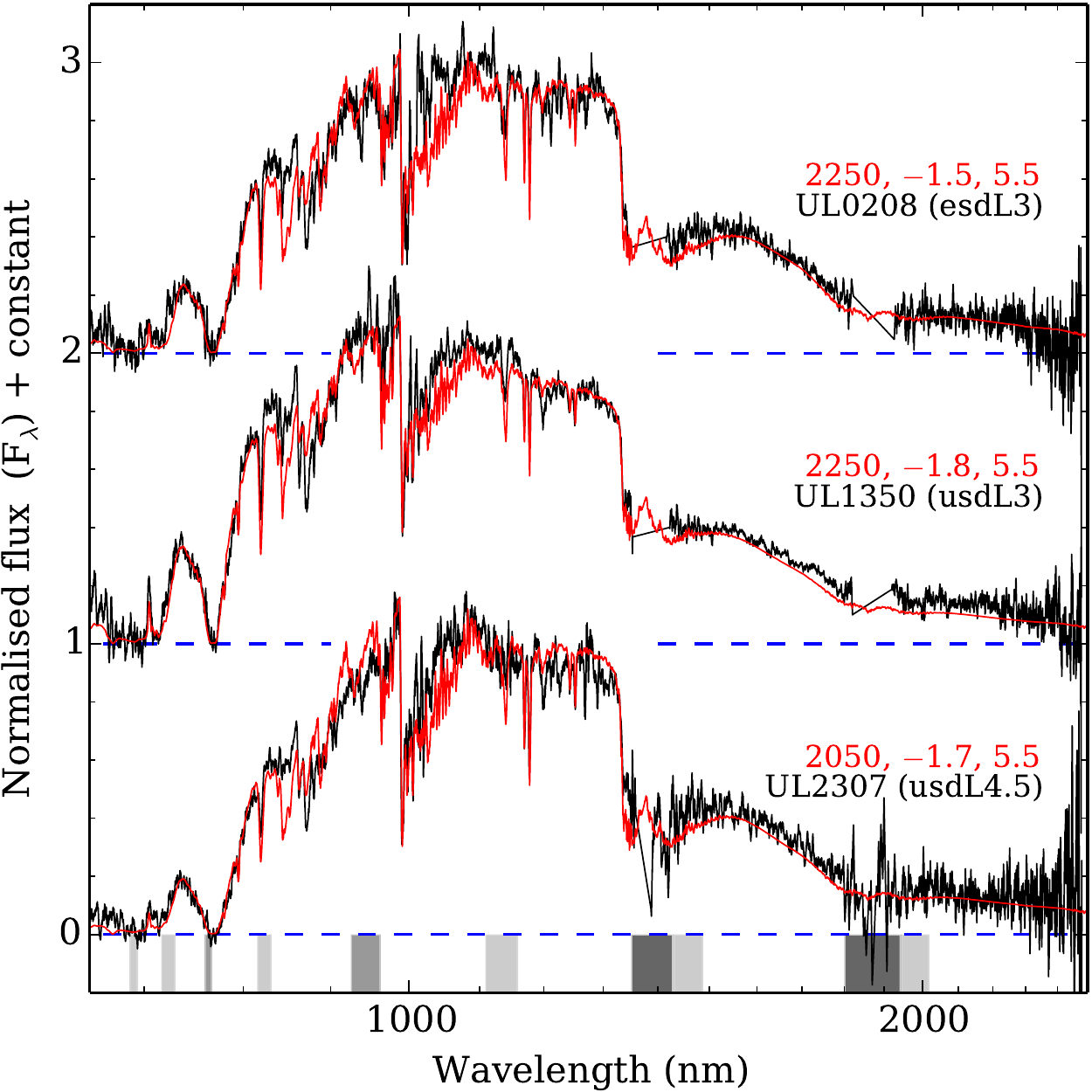}
\caption[]{The optical-NIR X-shooter spectra of UL0208, UL1350, and UL2307 obtained with the VLT, compared to their best fitting BT-Dusty model spectra. $T_{\rm eff}$, [Fe/H] and log $g$ values for each model are indicated next to the spectra. Spectra are normalised at 1100 nm. The axis tick marks are spaced logarithmically to more clearly display the features of the optical spectra.}
\label{mnir}
\end{center}
\end{figure}

Fig. \ref{mnir} shows the optical and NIR spectra of UL0208, Ul1350, and UL2307 compared to their best-fitting models. These best-fitting models have $T_{\rm eff}$ = 2250 K and [Fe/H] = $-$1.5 for UL0208;  $T_{\rm eff}$ = 2250 K and [Fe/H] = $-$1.8 for UL1350; $T_{\rm eff}$ = 2050 K and [Fe/H] = $-$1.7 for UL2307. All these models have gravity of log $g$ = 5.50 and $\alpha$ enhancement of [$\alpha$/Fe] = +0.4. These model spectra in general provide a very good fit to the data with the exception of the 840--880 nm region, where the TiO, CrH, and FeH absorption bands are less well fitted.

We have combined our new $T_{\rm eff}$ values with those from Paper I and Paper II, and extended this sample with an additional 50 K6.5--M5.5 subdwarfs in the $-2.0 \leq$ [Fe/H] $\leq -1.0$ range. Using this sample, we have updated the polynomial fit for the relationship between ultracool subdwarf SpT and $T_{\rm eff}$, which follows: 
\begin{equation}
\label{esptt}
T_{\rm eff} = 3701 - 108.6 \times SpT + 1.855 \times SpT^2 - 0.1661 \times SpT^3		
\end{equation}
with a root mean square (rms) of 32 K. In this equation SpT = $-$1 for K7, SpT = 0 for M0, SpT = 10 for L0 and SpT = 15 for L5 (etc). Equation \ref{esptt} is valid for K7--L7 subdwarfs with [Fe/H] $\la -1.0$.

\begin{figure}
\begin{center}
   \includegraphics[width=\columnwidth]{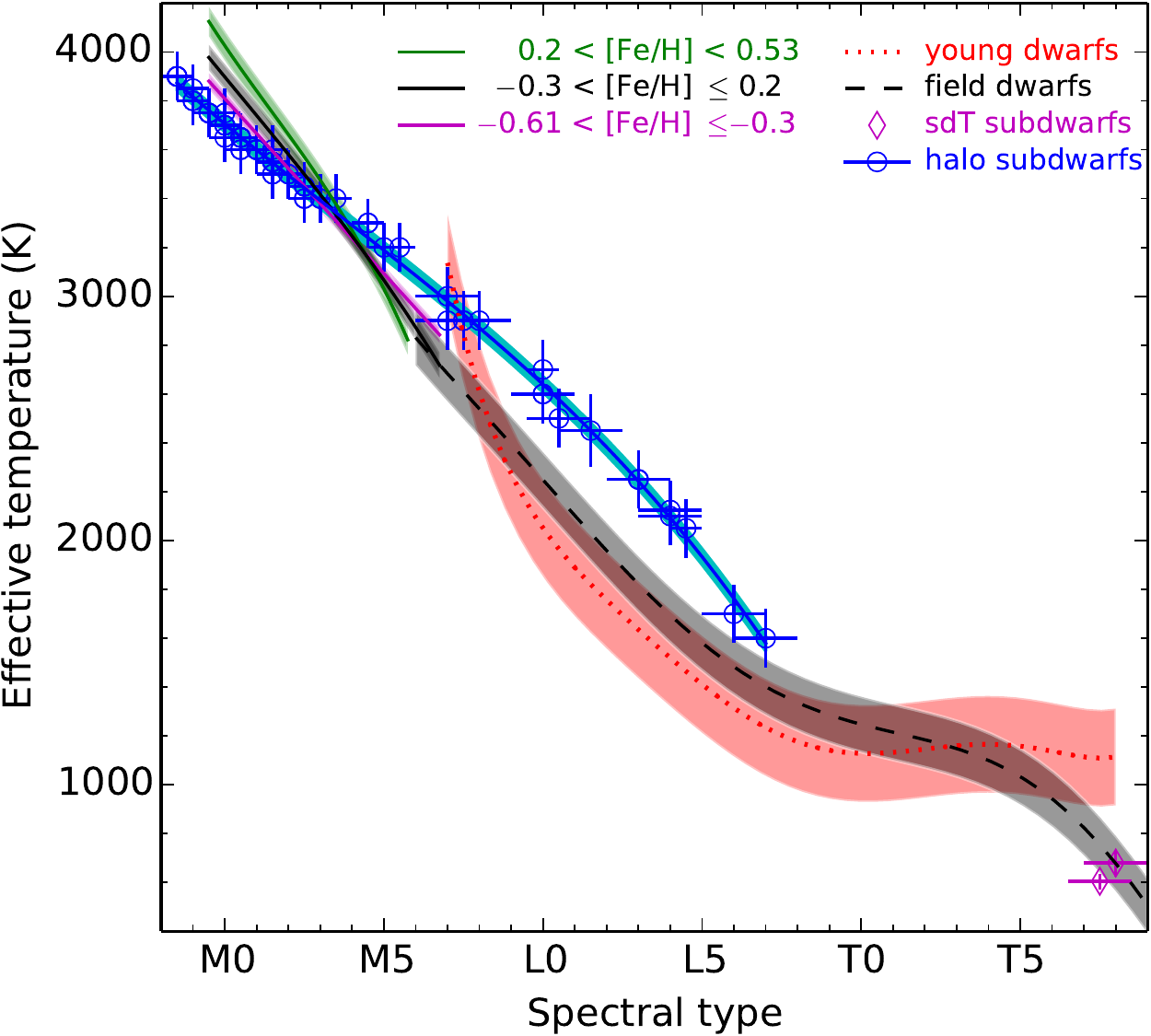}
\caption[]{The correlation between spectral type and $T_{\rm eff}$. The blue solid, black dashed, and red dotted lines are for halo subdwarfs, field dwarfs  \citep{fili15}, and young dwarfs \citep{fahe16}, respectively. The green, black and magenta solid lines are for M dwarfs in different metallicity ranges \citep{mann15}. Two magenta diamonds indicate two sdT subdwarfs: BD+01$\degr$ 2920 B \citep{pin12} and ULAS 1416+1348 \citep{line17}. }
\label{fsdmlt}
\end{center}
\end{figure}

Figure \ref{fsdmlt} shows the SpT--$T_{\rm eff}$ relationships of dwarf populations with different metallicities. Objects with lower metallicity tend to have lower $T_{\rm eff}$ before M3 and higher $T_{\rm eff}$ from M5 to L7 types. M6--L7 subdwarfs with [Fe/H] $\leq -1.0$ are $\sim$ 200--400 K warmer than field dwarfs. The SpT--$T_{\rm eff}$ relation for young L dwarfs is in general somewhat cooler (by $\sim$ 200 K) than the field relation, and is thus offset in the opposite sense to the subdwarf fit. The large difference between these SpT--$T_{\rm eff}$ relationships is due to different SpT--mass correlations, opacities, and ages of these populations. 
A metal-poor star would have a lower opacity, higher $T_{\rm eff}$, and bluer spectral continuum than a solar-metallicity star with the same mass, and thus could have the same spectral subtype as more massive stars with higher metallicities (fig. 9; Paper II).  Dwarf stars have higher mass than subdwarfs with the same spectral type before $\sim$ M6 and a sharper mass decline from M3 to M5 types \citep{mann15}. Then subdwarfs start to have higher masses from M7 to L types. Meanwhile, young ultracool dwarfs have much lower masses than field objects but retain a warm $T_{\rm eff}$ at young ages.

\begin{figure}
\begin{center}
   \includegraphics[width=\columnwidth]{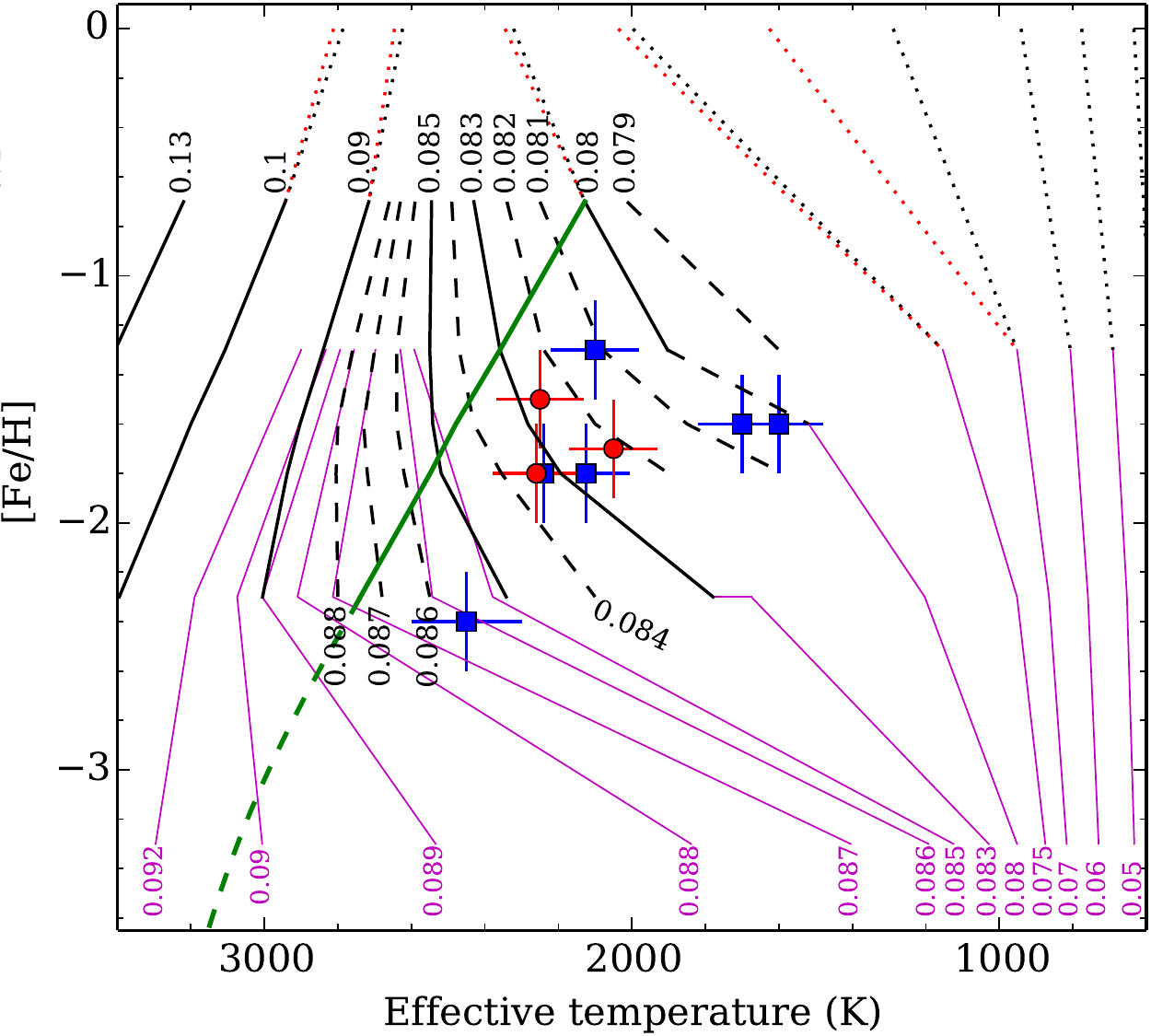}
\caption[]{Six known (blue squares) and three new (red circles) halo transitional brown dwarfs plotted in a [Fe/H] and $T_{\rm eff}$ space. Black solid and dashed (interpolated) lines indicate the 10 Gyr iso-mass contours derived from evolutionary models \citep{bara97,cha97}. Magenta solid lines are 10 Gyr iso-mass contours based on calculations of \citet{burr98}. The mass value (in M$_{\odot}$) is marked next to each iso-mass line. Note that the accuracy of these iso-mass contours is related to the density of the metallicity grid. The green line indicates the stellar--substellar boundary (Paper II). $T_{\rm eff}$ of solar-metallicity objects at 10 Gyr predicated by \citet{bara03,bara15} are joined with those of metal-poor models (with corresponding mass) by black and red dotted lines, respectively. These black dotted lines above [Fe/H] $= -0.6$ are not representing the true $T_{\rm eff}$ of thin- and thick disc brown dwarfs with corresponding masses, because stars with [Fe/H] $\ga -0.6$ usually have age $\la$ 8 Gyr \citep{red06}. }
\label{hbmm}
\end{center}
\end{figure}

\begin{table*}
 \centering
  \caption[]{Properties of nine known halo transitional brown dwarfs. SpT$_d$ and Ref$_d$ are the initial spectral type and reference. SpT$_a$ and Ref$_a$ are adopted spectral type (based on Paper I) and reference. Uncertainties on $T_{\rm eff}$, [Fe/H], and Mass ($M$) are approximately 120 K, 0.2 dex, and 0.0015 M$_{\odot}$, respectively.}
\label{thbd}
  \begin{tabular}{c cccccccc }
\hline
    Name  & ~SpT$_d$  & Ref$_d$ & SpT$_a$ & Ref$_a$ & $T_{\rm eff}$ (K) & [Fe/H] & $M$ (M$_{\odot}$)  \\
\hline
SDSS J010448.46+153501.8 & sdM9.5 & \citet{lod17} & usdL1.5 & Paper II &  2450 & $-$2.4 & 0.0860 \\
ULAS J020858.62+020657.0 & --- & --- & esdL3 & This paper & 2250 & $-$1.5 & 0.0827\\
2MASS J05325346+8246465 & sdL7 & \citet{bur03}  & esdL7 & Paper I & 1600 & $-$1.6 & 0.0802 \\
2MASS J06164006$-$6407194 & sdL5 & \citet{cus09}  & esdL6 & Paper I & 1700 & $-$1.6  & 0.0805\\
SDSS J125637.13$-$022452.4  & sdL4 & \citet{siv09}  & usdL3 & Paper I  & 2250 & $-$1.8 & 0.0833\\
ULAS J135058.86+081506.8  & sdL3.5--4 & \citet{lod17}  & usdL3 & This paper & 2250 & $-$1.8 &  0.0833\\
ULAS J151913.03$-$000030.0  & esdL4 & Paper I  & esdL4 & Paper I  & 2100 & $-$1.3  & 0.0811\\
2MASS J16262034+3925190  & sdL4  & \citet{bur04} & usdL4 & Paper I & 2125 & $-$1.8 & 0.0828 \\
ULAS J230711.01+014447.1 & --- & ---  & usdL4.5 & This paper  & 2050 & $-$1.7  & 0.0822 \\
\hline
\end{tabular}
\end{table*}

\subsection{Halo transitional brown dwarfs}
\label{shbd}
Evolutionary models \citep{bara97,cha97,burr98} show that there is a halo brown dwarf transition zone in a $T_{\rm eff}$ versus [Fe/H] plane, which is between $\sim$1200 K and the corresponding $T_{\rm eff}$ ($\sim$ 2200--3000 K) of the HBMM at different metallicities (Paper II). Six L subdwarfs have previously been identified as members of this transition zone. They are supported by electron degeneracy and unsteady nuclear ignition takes place in their cores, but they cannot balance ongoing gravitational contraction.  These halo transitional brown dwarfs are fully convective and could sporadically reach states of minimum temperature and pressure in their cores to fuse hydrogen. The unsteady hydrogen fusion in halo transitional brown dwarfs is less efficient and could last longer than the steady hydrogen fusion in least-massive stars which would have a main sequence lifetime of $\sim$ 12 trillion years  \citep{adam05}.

The unsteady hydrogen fusion partially contributed to the luminosities of halo transitional brown dwarfs. \citet{bur08} show that around half of the luminosity of 2MASS J05325346+8246465 is generated from its core hydrogen fusion at 10 Gyr. The efficiency (intensity or frequency) of the unsteady sporadic hydrogen fusion is sensitive to the masses of transitional brown dwarfs, because the halo brown dwarf transition zone covers a narrow mass range ($\sim$ 90--99\% HBMM) but spans wide ranges of luminosity and temperature  \citep[see figs 4 and 5 in][]{burr01}. For example, an object with 0.08 M$_{\odot}$ and [Fe/H] = $-$2.3 would have a $T_{\rm eff} \approx 1200$ K; however, an object with 0.085 M$_{\odot}$ and the same [Fe/H] would have a $T_{\rm eff} \approx 2400$ K (see Fig. \ref{hbmm}). 

Figure \ref{hbmm} shows UL0208, UL1350, and UL2307 in the $T_{\rm eff}$ versus [Fe/H] plane. They lie in the halo brown dwarf transition zone joined by six previously known members, which were identified using the method described in Paper II with atmospheric properties from Paper I. 
Masses of these nine halo transitional objects have been estimated from their location relative to the iso-mass contours in the diagram, and are shown in Table \ref{thbd}. The lowest-mass stars have stable $T_{\rm eff}$ at 10--13 Gyr \citep[e.g. fig 8 of][]{burr01}, so the iso-mass contours and the $T_{\rm eff}$ values of the HBMM line in Figure \ref{hbmm} are age-independent for halo stars. However, age uncertainty may affect our mass estimates for the halo brown dwarfs by up to 0.0005 M$_{\odot}$, because massive brown dwarf $T_{\rm eff}$ drops by $\sim$ 50--100 K from 10 Gyr to 11--13 Gyr \citep[e.g.][]{bara03}. The mass uncertainty caused by $T_{\rm eff}$ error is around 0.001 M$_{\odot}$, and the mass uncertainty caused by [Fe/H] error (0.2 dex) is around 0.008--0.001 M$_{\odot}$. The total uncertainty based on combining in quadrature is thus 0.0015 M$_{\odot}$. The mass uncertainty is thus small, and is mainly due to $T_{\rm eff}$ uncertainty (since $T_{\rm eff}$ is very sensitive to mass in the halo brown dwarf transition zone - a small mass change leads to a significant change in core fusion intensity).

An sdL7 subdwarf, ULAS J133836.97$-$022910.7 (UL1338), has $T_{\rm eff}$ = 1650$\pm120$ K and [Fe/H] = $-$1.0$\pm$0.2 (Paper I), and is a brown dwarf according to fig 9 of Paper II. However, UL1338 could belong to either the thick disc or the halo population as its [Fe/H] is at the thick-disc/halo boundary. Furthermore, its [Fe/H] could be underestimated, since its NIR spectrum has low SNR and the [Fe/H] derived from the BT-Settl model fits is less reliable for [Fe/H] $\geq -$1.0 than that for [Fe/H] < $-$1.0. Therefore, UL1338 currently is not listed as a halo transitional brown dwarf.  VVV J12564163-6202039 (VVV1256) is a high proper motion L subdwarf identified by \citet{smit18}. It could be either an sdL7 brown dwarf or an esdL1 star by its poor-quality optical-NIR spectrum and photometric colours. Better-quality optical-NIR spectra are required to confirm the status of VVV1256 and UL1338.

\subsection{Lithium detection in halo brown dwarfs}
Brown dwarfs with solar metallicity and mass less than $\sim$ 0.06 M$_{\odot}$ do not acquire core temperatures to fuse lithium ($\ga$ 2.5 million K), and retain this element throughout their lives \citep[e.g. fig 7 of][]{burr01}. The lithium depletion test \citep{rebo92,maga93} is a robust method to confirm young or field degenerate brown dwarfs with mass below 0.06 M$_{\odot}$ \citep[e.g.][]{basr96,rebo96,ruiz97,lod15b}. However, this method may be incapable of confirming old halo brown dwarfs. 

The maximum core temperature of a brown dwarf with 0.06 M$_{\odot}$ and [M/H] $\leq -$1.0 is almost 0.2 million K below the minimum temperature to fuse lithium \citep[fig. 10;][]{cha97}. A core temperature difference of 0.2 million K is corresponding to a mass difference of $\sim$ 0.005 M$_{\odot}$ for brown dwarfs in this mass range \citep[fig. 2;][]{burr01}. Therefore, the lithium-burning minimum mass is likely around 0.065 M$_{\odot}$ at [M/H] $\leq -$1.0. The Li line at 670.78 nm was not detected in the spectra of SD1256 and 2M1626 \citep{lod15a}. Note that their masses are $\sim$ 27\% higher than 0.065 M$_{\odot}$ (see Table \ref{thbd}). Halo brown dwarfs with [M/H] $\leq -$1.0 (i.e. [Fe/H] $\leq -$1.3) and masses below 0.065 M$_{\odot}$ would have $T_{\rm eff} \la$ 900 K at 10 Gyr, according to Fig. \ref{hbmm}. 
Brown dwarfs with $T_{\rm eff} \la$ 900 K would have spectral types later than T5 (Fig. \ref{fsdmlt}) and barely have flux at $<$ 770 nm to unveil monatomic lines.

Lithium starts to condense into LiCl as well as LiOH and LiH in the atmospheres of brown dwarfs when temperatures drop down below $\sim$ 1525 K, and monatomic lithium decreases in abundance with decreasing temperature \citep{lodd99}.  Therefore, Li lines are not expected in the optical spectra of halo brown dwarfs, which could be either massive enough to fuse lithium to helium or cool enough to condense lithium into LiCl, LiOH, or LiH.  

Metal hydrides (e.g. FeH) are favoured in the atmospheres of metal-poor L subdwarfs (Paper I). \citet{shi13} also suggest that LiH might be a better candidate for detection of lithium molecules in the mid-infrared spectra of brown dwarfs than LiCl \citep{weck04}. Therefore, LiH might be more important than LiCl and LiOH in halo degenerate brown dwarfs, which would have insufficient metals (e.g. Cl, O) to combine with lithium.

\section{Summary and conclusions}
\label{ssac}

We have presented optical and NIR spectra of two new L subdwarfs (UL0208, UL2307) and a known L subdwarf (UL1350) obtained with OSIRIS and X-shooter. We classified them as esdL3, usdL4.5, and usdL3, respectively. We measured their astrometry and kinematics, and determined their $T_{\rm eff}$ and [Fe/H] by fitting model spectra. We determined their mass by comparing to model iso-mass contours in the $T_{\rm eff}$ versus [Fe/H] plane and found that these three objects all have masses below the HBMMs at their metallicities. They therefore join a population of nine halo transitional brown dwarfs known to date, which we summarize in Table \ref{thbd}. 

 The unsteady core hydrogen fusion in halo transitional brown dwarfs partially contributed to their luminosities and helped them to maintain intermediate luminosities and $T_{\rm eff}$ between least-massive stars and degenerate brown dwarfs. Halo transitional brown dwarfs are distributed in a  $T_{\rm eff}$ range wider than 1000 K within a narrow mass range and naturally form a `substellar subdwarf gap'. Objects in this gap have a maximum mass relative difference of $\sim$ 10\% and a maximum $T_{\rm eff}$ relative difference of $\sim$ 2.5 times.  
Note that a substellar gap would not be distinctly revealed by luminosity or $T_{\rm eff}$ among field dwarf populations, as field transitional brown dwarfs have a relatively shorter cooling time (narrower $T_{\rm eff}$ range) and a higher level of mass/age degeneracy. The substellar gap is revealed in the simulated substellar luminosity function with a halo-like birthrate but not in that with a flat birthrate \citep[fig. 10;][]{burg04}. 

It is important to discover more of these halo brown dwarfs. Deep exploration of the Visible and Infrared Survey Telescope for Astronomy \citep[VISTA;][]{suth15} surveys and the future {\sl Euclid} survey \citep{laur11} would allow us to identify cooler halo brown dwarfs. The {\sl Euclid} space mission will be particularly well suited to identifying halo brown dwarfs because it will provide NIR slit-less spectra, in addition to its deep optical and $YJH$-band wide-angle imaging. The growing population of known transition-zone brown dwarfs can be used as a benchmark to search for halo brown dwarfs in {\sl Euclid}. Meanwhile, more advanced atmospheric and evolutionary models are required to better understand cool halo brown dwarfs.

\section*{Acknowledgements}
Based on observations made with the Gran Telescopio Canarias (GTC), installed in the Spanish Observatorio del Roque de los Muchachos of the Instituto de Astrof{\'i}sica de Canarias, on the island of La Palma. Based on observations collected at the European Organisation for Astronomical Research in the Southern Hemisphere under ESO programmes 094.C-0202, 095.C-0878, 096.C-0130, and 096.C-0974. This work is based in part on data obtained as part of the UKIRT Infrared Deep Sky Survey. The UKIDSS project is defined in \citet{law07}. UKIDSS uses the UKIRT Wide Field Camera \citep[WFCAM;][]{casa07}. The photometric system is described in \citet{hew06}, and the calibration is described in \citet{hodg09}. The pipeline processing and science archive are described in \citet{irwi04} and \citet{hamb08}. Funding for the Sloan Digital Sky Survey (SDSS) has been provided by the Alfred P. Sloan Foundation, the Participating Institutions, the National Aeronautics and Space Administration, the National Science Foundation, the U.S. Department of Energy, the Japanese Monbukagakusho, and the Max Planck Society. The SDSS website is \url{http://www.sdss.org/}. 
This publication makes use of data products from the Pan-STARRS1 Surveys. 
This research has benefited from the SpeX Prism Spectral Libraries, maintained by Adam Burgasser at \url{http://www.browndwarfs.org/spexprism}. 

ZHZ is supported by the PSL fellowship. ZHZ acknowledges the financial support of the National Fundamental Research Program of China (973 Program, 2014 CB845705). DJP and HRAJ acknowledge support from the UK's Science and Technology Facilities Council, grant numbers ST/M001008/1, ST/N001818/1, and ST/J001333/1. NL is funded by the Spanish Ministry of Economy and Competitiveness (MINECO) under the grants AYA2015-69350-C3-2-P. EM is funded by the MINECO under grant AYA2015-69350-C3-1-P. MCGO acknowledges the financial support of a JAE-Doc CSIC fellowship co-funded with the European Social Fund under the programme Junta para la Ampliaci{\'o}n de Estudios and the support of the Spanish Ministry of Economy and Competitiveness (MINECO) through the project AYA2014-54348-C3-2-R.
DH is supported by Sonderforschungsbereich SFB 881 `The Milky Way System' (subproject A4) of the German Research Foundation (DFG). FA received funding from the French `Programme National de Physique Stellaire' (PNPS) and the `Programme National de Plan\'etologie' of CNRS (INSU). The computations of atmosphere models were performed  in part on the Milky Way supercomputer, which is funded by the Deutsche Forschungsgemeinschaft (DFG) through the Collaborative Research Centre (SFB 881) `The Milky Way System' (subproject Z2) and hosted at the University of Heidelberg Computing Centre, and at the {\sl P\^ole Scientifique de Mod\'elisation Num\'erique} (PSMN) at the {\sl \'Ecole Normale Sup\'erieure} (ENS) in Lyon, and at the {\sl Gesellschaft f{\"u}r Wissenschaftliche Datenverarbeitung G{\"o}ttingen} in collaboration with the Institut f{\"u}r Astrophysik G{\"o}ttingen. The authors thank the referee for the useful and constructive comments.

\end{document}